\documentclass[a4paper]{jpconf}
\usepackage{graphicx}
\usepackage{iopams}
\begin{document}
\title{Kalman filter tracking on parallel architectures}

\author{G Cerati$^{1,}\footnote[4]{Present address: Fermi National Accelerator Laboratory, Batavia, USA}$,
P Elmer$^2$, S Krutelyov$^1$, S Lantz$^3$, M Lefebvre$^2$, K McDermott$^3$, D Riley$^3$, M Tadel$^1$, P Wittich$^3$, F Würthwein$^1$, and A Yagil$^1$}

\address{$^1$ UC San Diego, 9500 Gilman Dr., La Jolla, California, USA 92093}
\address{$^2$ Princeton University, Princeton, New Jersey, USA 08544}
\address{$^3$ Cornell University, Ithaca, New York, USA}

\ead{Daniel.Riley@cornell.edu}

\begin{abstract}
% Limits on power dissipation have pushed CPUs to grow in parallel processing capabilities rather than clock rate, leading to the rise of ``manycore'' or GPU-like processors. In order to achieve the best performance, applications must be able to take full advantage of vector units across multiple cores, or some analogous arrangement on an accelerator card. Such parallel performance is becoming a critical requirement for methods to reconstruct the tracks of charged particles at the Large Hadron Collider and, in the future, at the High Luminosity LHC. This is because the steady increase in luminosity is causing an exponential growth in the overall event reconstruction time, and tracking is by far the most demanding task for both online and offline processing. Many past and present collider experiments adopted Kalman filter-based algorithms for tracking because of their robustness and their excellent physics performance, especially for solid state detectors where material interactions play a significant role. 
We report on the progress of our studies towards a Kalman filter track reconstruction algorithm with optimal performance on manycore architectures. The combinatorial structure of these algorithms is not immediately compatible with an efficient SIMD (or SIMT) implementation; the challenge for us is to recast the existing software so it can readily generate hundreds of shared-memory threads that exploit the underlying instruction set of modern processors. We show how the data and associated tasks can be organized in a way that is conducive to both multithreading and vectorization. We demonstrate very good performance on Intel Xeon and Xeon Phi architectures, as well as promising first results on Nvidia GPUs.
% We discuss the current limitations and the plan to achieve full scalability and efficiency in collision data processing.
\end{abstract}

\section{Introduction}

\begin{figure}[htb]
  \begin{center}
   \includegraphics[width=0.5\textwidth]{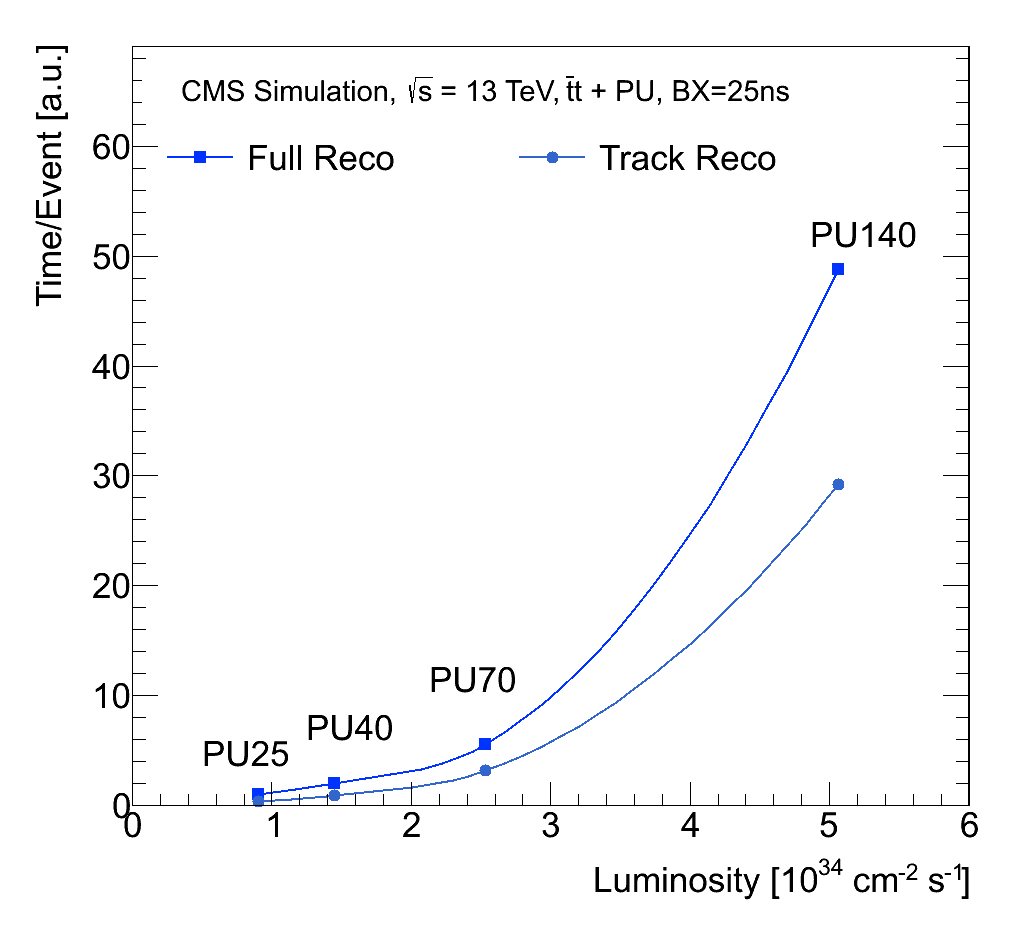}
   \caption{CPU time per event versus instantaneous luminosity, for both full reconstruction and the dominant tracking portion. PU25 corresponds to the data taken during 2012, and PU140 corresponds to the low end of estimates for the HL-LHC era..}  
   \label{fig:eff_tracking_pileup}
  \end{center}
\end{figure}

Of the steps reconstructing events in the CMS detector, tracking is the most computationally complex, most sensitive to activity in the detector, and least amenable to parallelization. The speed of online reconstruction has a direct impact on how effectively interesting data can be selected from the 40 MHz collisions rate, while the speed of the offline reconstruction limits how much data can be processed for physics analyses.  The large time spent in tracking will become even more important in the HL-LHC era of the Large Hadron Collider. The increase in event rate will increase the detector occupancy (“pile-up”, PU), leading to the exponential gain in time taken to perform track reconstruction illustrated in Figure~\ref{fig:eff_tracking_pileup}\cite{pileup}.

At the same time, due to power density and physical scaling limits, the performance of single CPUs has slowed, with advances in performance increasingly relying on multi-core or highly parallel architectures.  In order to sustain the higher HL-LHC processing requirements without compromising physics performance and timeliness of results, the LHC experiments must make full use of highly parallel architectures.  As a potential solution, we are investigating adapting the existing CMS track finding algorithms and data structures\cite{cmstrack} to make efficient use of highly parallel architectures, such as Intel’s Xeon Phi and Nvidia GPUs.  In this paper we provide an update to results most recently reported at CHEP2015\cite{cerati-chep15} and Connecting the Dots 2016\cite{ctd-2016}, including our first results using  Intel Threaded Building Blocks\cite{tbb} (TBB) instead of OpenMP for multi-threading support.

\section{Kalman Filter Tracking}

Our targets for parallel processing are track reconstruction and fitting algorithms based on the Kalman Filter\cite{kalman} (KF). KF-based tracking algorithms are widely used to incorporate estimates of multiple scattering directly into the trajectory of the particle. Other algorithms, such as Hough Transforms\cite{hough1}\cite{hough2} and Cellular Automata\cite{cellular1}\cite{cellular3} adapted from image processing applications, are more naturally parallelized.  However, these are not the main algorithms in use at the LHC today. The LHC experiments have an extensive understanding of the physics performance of KF algorithms; they have proven to be robust and perform well in the difficult experimental environment of the LHC.

KF tracking proceeds in three main stages: seeding, building, and fitting. Seeding provides the initial estimate of the track parameters based on a few hits in the innermost regions of the detector. Realistic seeding is currently under development and will not be reported here. Track building projects the track candidate outwards to collects additional hits, using the KF to estimate which hits represent the most likely continuation of the track candidate. Track building is by far the most time consuming step of tracking, as it requires branching to explore multiple candidate tracks per seed after finding compatible hits on a given layer. When a complete track has been reconstructed, a final fit using the KF is performed to provide the best estimate of the track's parameters.

To take full advantage of parallel architectures, we need to exploit two types of parallelism: vectorization and parallelization. Vector operations perform a single instruction on multiple data at the same time, in lockstep. In tracking, branching to explore multiple candidates per seed can interfere with lock-step single-instruction synchronization. Multi-thread parallelization aims to perform different instructions at the same time on different data. The challenge to tracking is workload balancing across different threads, as track occupancy in a detector is not uniformly distributed on a per event basis. Past work by our group has shown progress in porting sub-stages of KF tracking to support parallelism in simplified detectors (see, e.g. our presentations at ACAT2014\cite{acat2014} and CHEP2015\cite{cerati-chep15}). As the hit collection is completely determined after track building, track fitting can repeatedly apply the KF algorithm without branching, making this a simpler case for both vectorization and parallelization; first results in porting KF tracking to Xeon and Xeon Phi were shown at ACAT2014\cite{acat2014}.

\subsection{Matriplex}

The implementation of a KF-based tracking algorithm consists of a sequence of operations on matrices of sizes from $3\times 3$ up to $6\times 6$.  In order to optimize efficient vector operations on small matrices, and to decouple the computational details from the high level algorithm, we have developed a new matrix library. The {\it Matriplex} memory layout uses a matrix-major representation optimized for loading vector registers for SIMD operations on small matrices, using the native vector-unit width on processors with small vector units or very large vector widths on GPUs.  Matriplex includes a code generator for defining optimized matrix operations, with support for symmetric matrices and on-the-fly matrix transposition. Patterns of elements that are known by construction to be zero or one can be specified, and the resulting  code will be optimized to eliminate unnecessary register loads and arithmetic operations. The generated code can be either standard C++ or macros that map to architecture-specific intrinsic functions.

\subsection{Test Scenarios}\label{sec:scenarios}

In order to study parallelization with minimal distractions, we developed a standalone KF-based tracking code using a simplified ideal barrel geometry with a uniform longitudinal magnetic field, gaussian-smeared hit positions, a particle gun simulation with flat transverse momentum distribution between 0.5 and 10 GeV, no material interaction, and no correlation between particles nor decays.  This simplified configuration was used to study vector and parallel performance and to study the performance of different choices of coordinate system and handling of KF-updates.  These studies led to the use of a hybrid coordinate system, using global Cartesian coordinates for spatial positions and polar coordinates for the momentum vector.

Recently we have begun work on using a more realistic CMS detector geometry.  Event data from the full CMS simulation suite is translated into a simplified form that can be processed by our standalone tracking code.  For track building, track propagation is performed in two steps, an initial propagation to the average radius of the barrel or average Z of the endcap disk, followed by a second propagation to the exact hit position for each candidate hit.  KF updates are performed on the plane tangent to the hit radius.  These choices make it possible to use a simplified form of the CMS geometry, avoiding the complexities of the full detector description.

\subsection{Performance Tuning}\label{sec:tuning}

Achieving acceptable vector and multi-thread parallel performance requires careful attention to detail.  Regular profiling with Intel VTune Amplifier and attention to the compiler optimization reports helped identify many obstacles, some relatively subtle.  We found that references to unaligned locations in aligned data structures may force the compiler into unaligned accesses, reducing vector performance; prefetching, scatter/gather instructions and other intrinsics need to be used in an organized, systematic fashion; unwanted conversions from float to double can reduce effective vector width; variables should be declared in the smallest scope possible, and use ``const'' wherever possible.

\begin{figure}[htb]
  \begin{center}
   \includegraphics[width=0.5\textwidth]{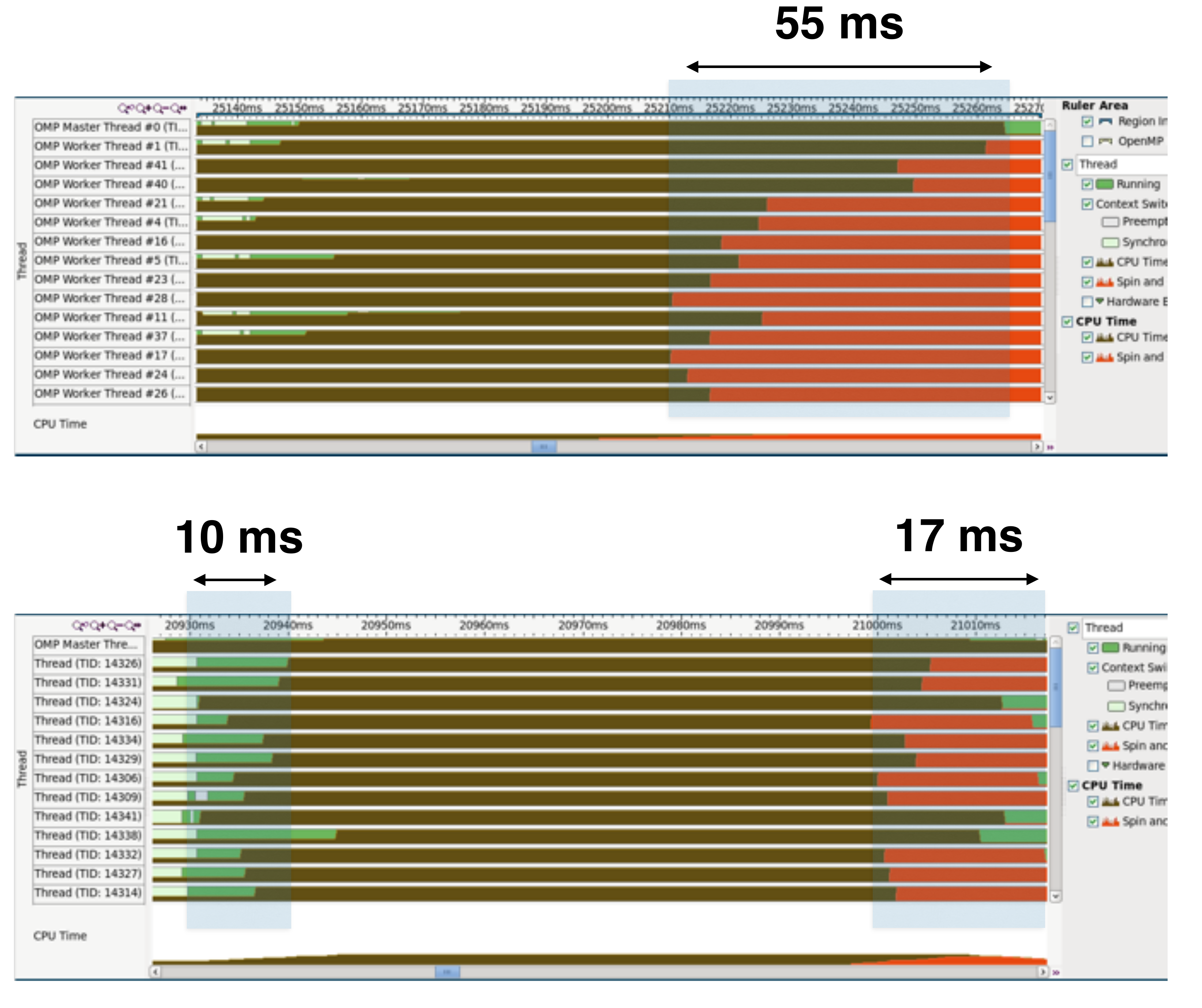}
   \caption{Comparison of thread occupancy with VTune Amplifier. The top figure shows thread occupancy with OpenMP and static workload assignment; the bottom with TBB and dynamic workload assignment shows smaller tail effects due to uneven workload.}
   \label{fig:tbb-vtune}
  \end{center}
\end{figure}

For multi-thread parallelism, the two most critical issues we found were memory management and workload balancing.  To avoid memory stalls and cache conflicts, we reduced our data structures to the minimum necessary for the algorithm, optimized our data structures for efficient vector operations, and minimized object instantiations and dynamic memory allocations.  Workload balancing for track building is complicated by the uncertain distribution of track candidates and hits, which can result in large tail effects for a naive static partitioning scheme.  We found that using Intel Threaded Building Blocks tasks, with smaller units of work and dynamic ``work-stealing'', let us naturally express thread-parallelism in a way that allowed more dynamic allocation of resources and reduced the tail effects from uneven workloads.  Figure~\ref{fig:tbb-vtune} illustrates the difference in tail effects between our initial OpenMP implementation with static workload partitioning compared to dynamic partitioning using TBB.

\section{Latest Results}

The primary platforms discussed are a Xeon Phi 7120P ``Knights Corner'' (KNC) processor and Xeon E5-2620 ``Sandy Bridge'' (SNB) system, with brief discussion of preliminary results on a Xeon Phi 72xx ``Knights Landing'' (KNL) processor and the Nvidia Tesla K20/K40 and Pascal P100.  Scaling on the more traditional Xeon architecture is in most cases near optimal, so the discussion primarily focus on the more highly parallel architectures.

\begin{figure}[htb]
  \begin{center}
   \includegraphics[width=\textwidth]{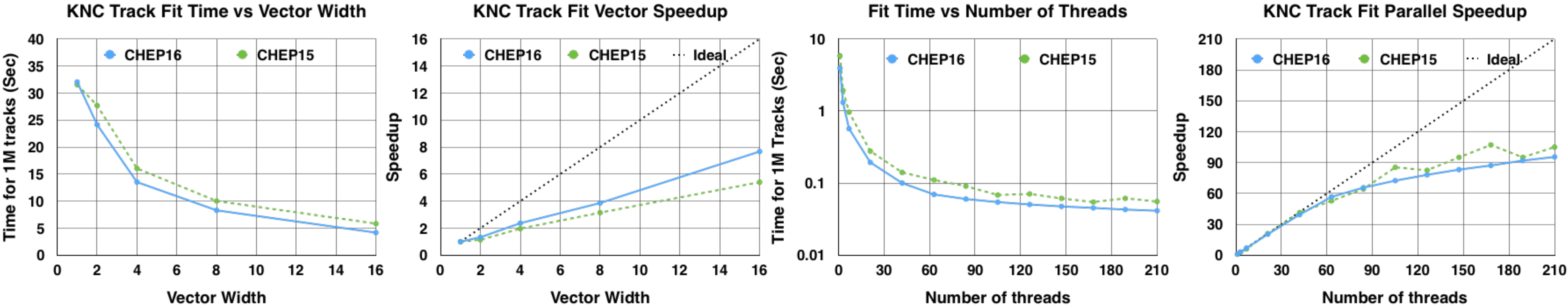}
   \caption{Comparison of CHEP2015 and CHEP2016 fitting performance.  The two graphs on the left show time and scaling as a function of vector width; the right two show vs. the number of threads.}
   \label{fig:fit-compare}
  \end{center}
\end{figure}

\subsection{Track Fitting}

Vector and multi-thread absolute performance and scaling improved mostly from our CHEP2015 results, with vectorization scaling showing the most improvement, shown in Figure~\ref{fig:fit-compare}.  This result is somewhat deceptive, as there are significantly more operations performed in the latest results due to the change of momentum to polar coordinates (from global Cartesian coordinates).  The change of coordinate system results in significantly more complicated off-diagonal terms in the propagation matrix.  The cost of these additional calculations was nearly exactly canceled by improvements to the memory and cache usage, and better vectorization.  On KNC, we achieve a vector speedup of nearly eight times, approximately half the theoretical maximum with a vector width of 16.  Multi-thread scaling is near ideal up to 61 threads, reaching nearly 100x speedup with 200 threads.  The KNC processor used has 61 physical cores, but must alternate between hyper-threads to fill all the available instruction slots, so ideal scaling would be a speedup by a factor of 122.  The ``knee'' in the scaling curve indicates that with two threads per core we are not achieving full utilization, possibly due to remaining memory bandwidth or cache effects.

\begin{figure}[htb]
  \begin{minipage}{14pc}
   \includegraphics[width=14pc]{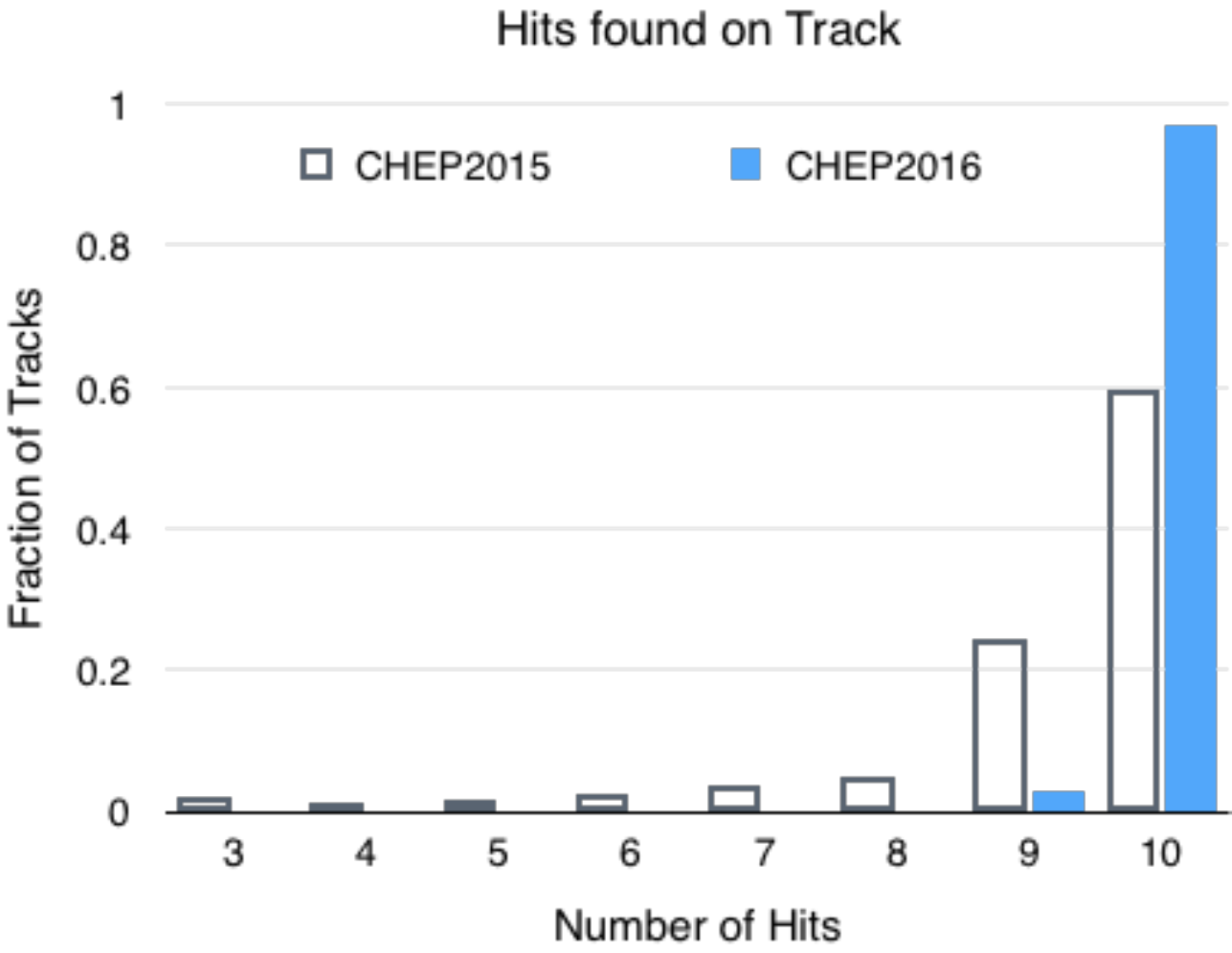}
  \end{minipage}\hspace{2pc}%
  \begin{minipage}{14pc}
   \includegraphics[width=14pc]{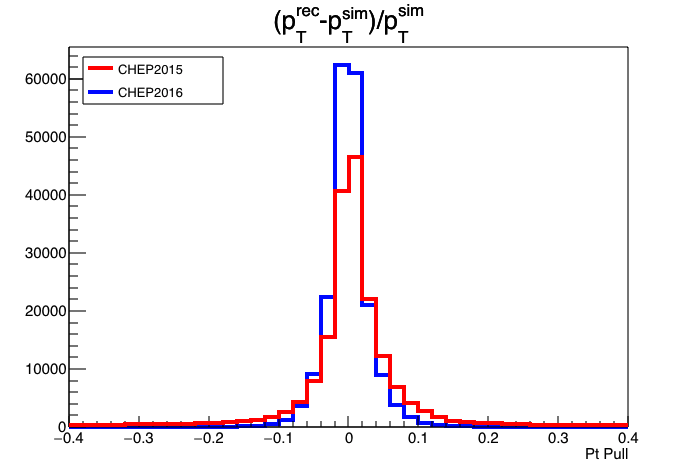}
  \end{minipage}
 \caption{Comparison of CHEP2015 and CHEP2016 track building physics performance.  The figure on the left shows fraction of tracks found vs. number of hits for a ten-layer detector; the figure on the right compares transverse momentum resolution.}
 \label{fig:physics-perf}
\end{figure}

\subsection{Track Building}

The change in coordinate system and other improvements significantly improved KF update stability, resulting in the much improved track parameter resolution and hit finding efficiency, shown in Figure~\ref{fig:physics-perf}.

\subsubsection{Vectorization}

\begin{figure}[htb]
  \begin{center}
   \includegraphics[width=\textwidth]{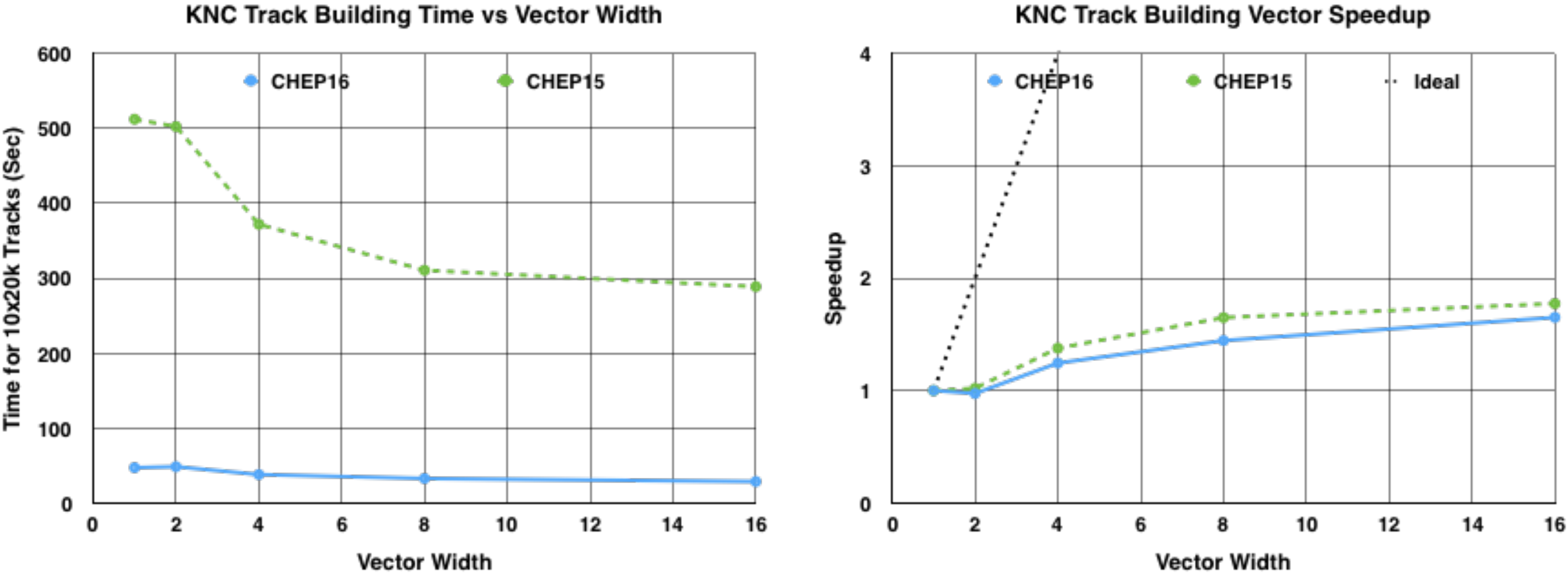}
   \caption{Comparison of CHEP2015 and CHEP2016 track building vector performance, showing time and scaling as a function of vector width.}
   \label{fig:build-vector}
  \end{center}
\end{figure}

The absolute time for single-thread track building was reduced by ~90\% due to the performance tuning discussed in Section~\ref{sec:tuning}.  However, effective vectorization of the track building remains a challenge, as shown in Figure~\ref{fig:build-vector}.  The combinatorial nature of the track building algorithm, which examines and adds a variable number of hit candidates in each layer, results in many branching operations that impede vectorization, as well as adding frequent repacking operations to keep the full vector width utilized.  The larger and more complicated data structures used for selecting hit candidates also results in poorer data locality and higher bandwidth requirements.  We are continuing to investigate possible strategies for improving the track building vector performance.

\subsubsection{Parallelization}

\begin{figure}[htb]
  \begin{center}
   \includegraphics[width=\textwidth]{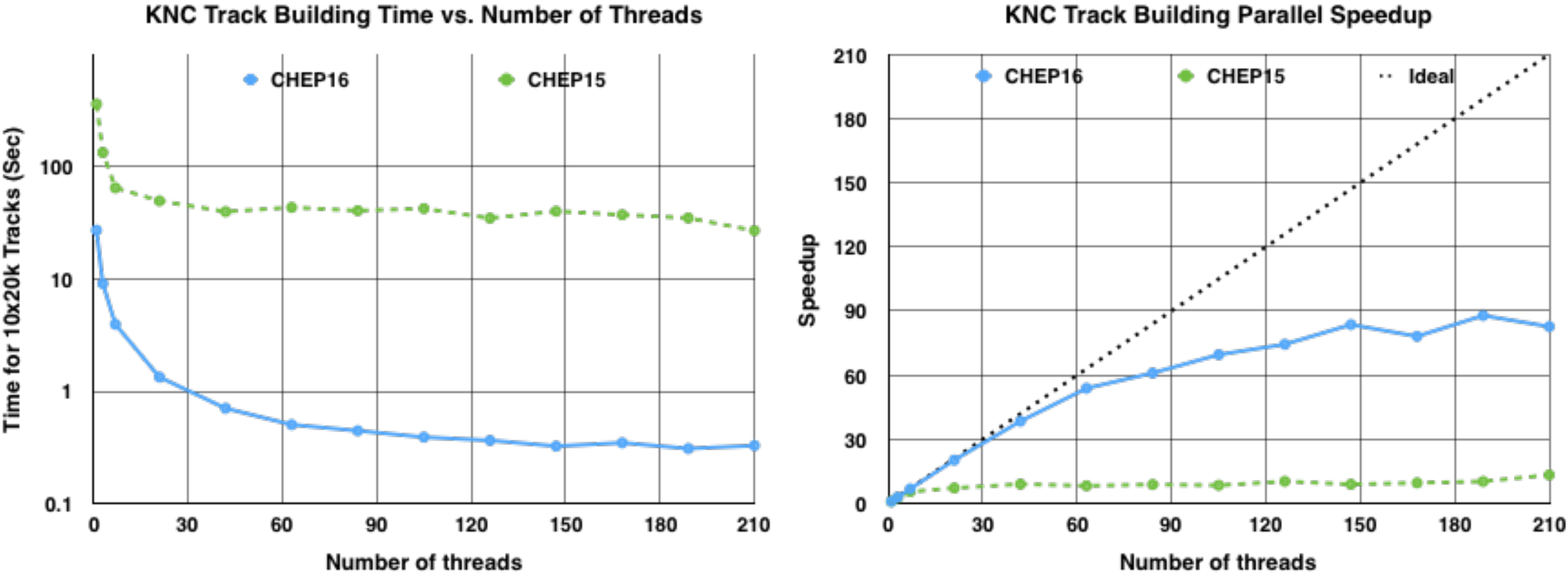}
   \caption{Comparison of CHEP2015 and CHEP2016 track building multi-thread performance, showing time and scaling as a function of the number of threads.}
   \label{fig:build-parallel}
  \end{center}
\end{figure}

Compared to our previous results, as shown in Figure~\ref{fig:build-parallel}, multi-thread track scaling has improved significantly, scaling within 15\% of ideal up to 61 threads and reaching approximately 80x speedup with 200 threads.  This result is very similar to the multi-thread scaling of the much simpler track fitting task.  Combining the improvements to the single-thread track finding efficiency and multi-thread scaling, track building with 200 threads is nearly 100 times faster than our CHEP2015 result.  These improvements illustrate the results of careful performance tuning combined with the more flexible workload balancing enabled by the switch to TBB for multi-threading.

\section{CMS Geometry}

\begin{figure}[htb]
  \begin{center}
   \includegraphics[width=\textwidth]{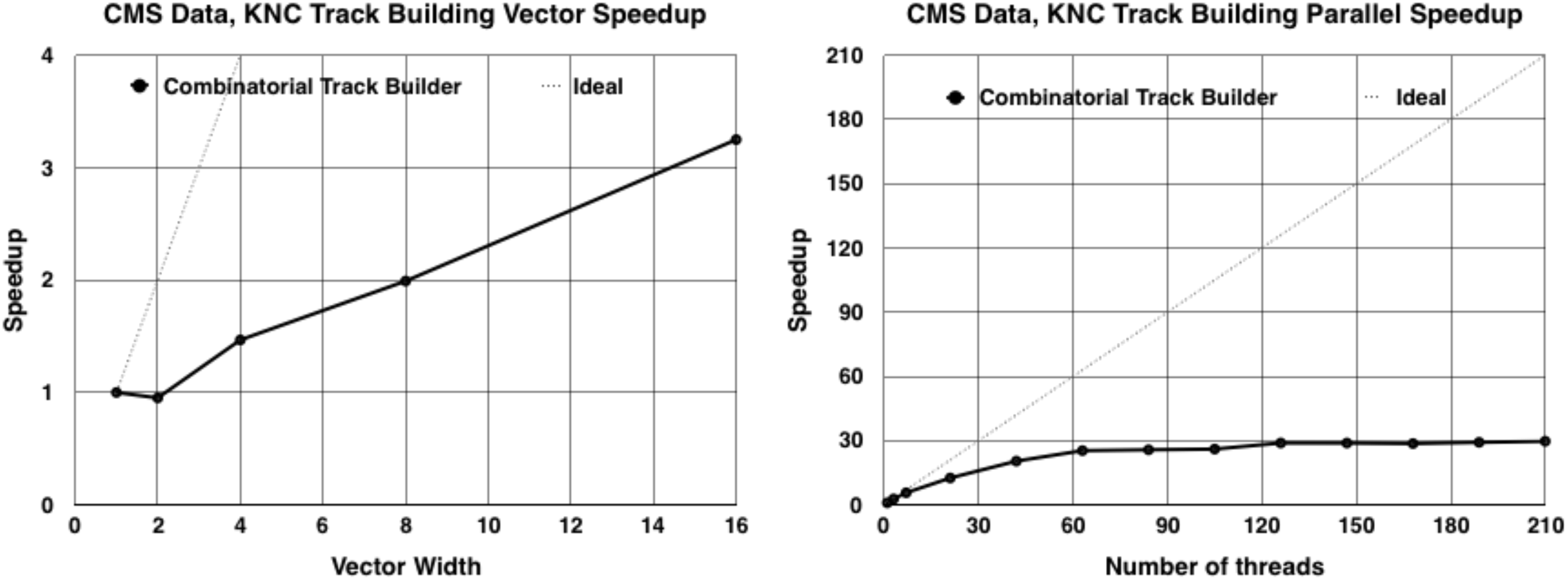}
   \caption{Scaling of track building with a simplified CMS geometry as a function of vector width (left) and number of threads (right).}
   \label{fig:cms-scaling}
  \end{center}
\end{figure}

Preliminary results for track building using the simplified CMS geometry discussed in Section~\ref{sec:scenarios} are shown in Figure~\ref{fig:cms-scaling}.  These results show better vector scaling, which we believe is due to the two-step propagation, first to the average radius and then to the hit radius.  The propagation routine vectorizes well, so the additional computation results in more time spent in routines with good vector register utilization.  Multi-thread scaling is significantly worse than our ``toy'' setup.  For this test we use seeds from the first step of the CMS iterative tracker, which yields around 500 seed tracks per event, compared to 20,000 Monte Carlo ``truth'' seeds in our toy tests.  Repeating our tests with the idealized scenario with 500 tracks per event shows scaling similar to the CMS geometry tests.  This is not surprising, as with only one event being processed at a time we need $200 \times 16 = 3200$ seeds to fill the vector registers of 200 threads.  We are currently working on processing multiple events at a time in a way that allows us to use seeds from multiple events in the same Matriplex so that we can fully use the vector registers.

\section{KNL and GPU Preliminary Results}

\begin{figure}[htb]
  \begin{minipage}{16pc}
   \includegraphics[width=\textwidth]{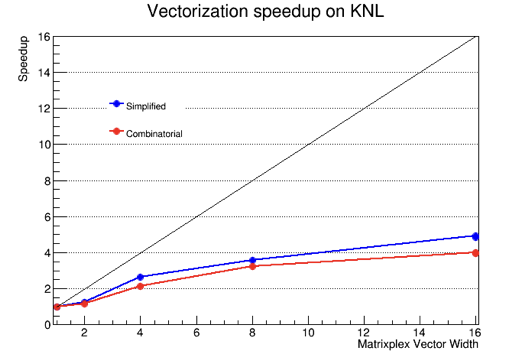}
  \end{minipage}\hspace{2pc}%
  \begin{minipage}{16pc}
   \includegraphics[width=\textwidth]{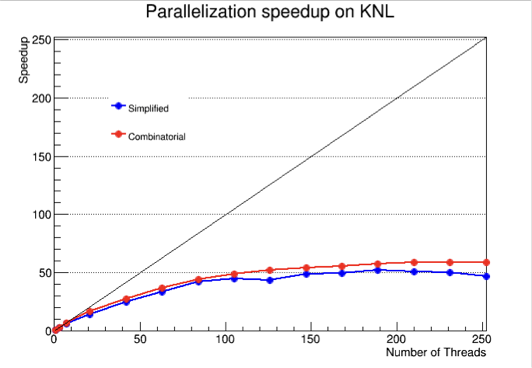}
  \end{minipage}
 \caption{Scaling of track building on KNL as a function of vector width (left) and number of threads (right).}
 \label{fig:knl-perf}
\end{figure}

We have run very preliminary, untuned tests of our track building on a KNL system, with the results shown in Figure~\ref{fig:knl-perf}.  On KNL we see better vectorization than KNC, and similar multi-thread scaling.  We find it encouraging that our track building algorithms show decent scaling performance on SNB, KNC and KNL without significant platform-specific scaling beyond simply matching the platforms vector register width.

\begin{figure}[htb]
  \begin{center}
   \includegraphics[width=0.6\textwidth]{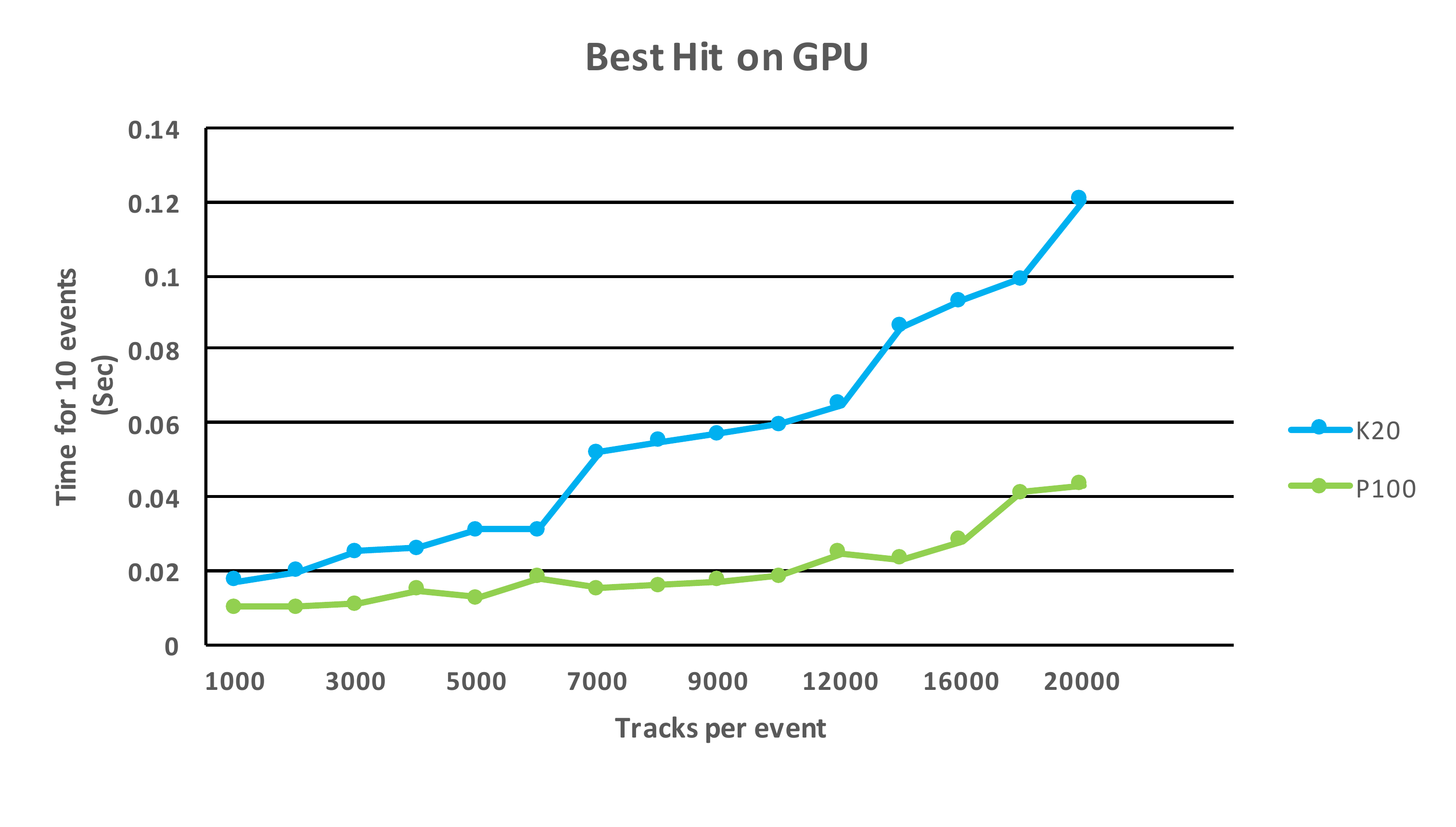}
   \caption{Comparison of K20 and Pascal P100 track building performance .}
   \label{fig:gpu-perf}
  \end{center}
\end{figure}

We have also implemented the track fitting algorithms and a simpler (non-combinatoric) version of the track building on Nvidia GPUs using the CUDA toolkit.  The GPU version uses a templated GPlex structure that matches the interfaces and layout of the Matriplex class, allowing substantial code sharing of the KF routines, while the higher-level ``steering'' routines are somewhat different due to different setup and memory management requirements of the different platforms.  The GPU implementations show good scaling of the KF routines but the total time tends to be dominated by setup times copying data structures to the GPU.  Preliminary tests with a Pascal P100 GPU, shown in Figure~\ref{fig:gpu-perf}, show better scaling with simpler memory management.

\section{Conclusion and Outlook}

We have made significant progress in parallelized and vectorized Kalman Filter-based end-to-end tracking R\&D on Xeon and Xeon Phi architectures, with some initial work on GPUs. Through the use of a variety of tools we have developed a good understanding of bottlenecks and limitations of our implementation which has led to further improvements. Through the use of our own Matriplex package and Intel Threaded Building Blocks (TBB), we can achieve good utilization of unconventional highly parallel vector architectures.  We are currently focusing on  processing fully realistic data, with encouraging preliminary results.

\ack

This work is supported by the U.S. National Science Foundation, under the grants PHY-1520969, PHY-1521042, PHY-1520942 and PHY-1120138, and by the U.S. Department of Energy.

\end{document}